\documentclass[twocolumn]{emulateapj}

\usepackage{amsmath}

\newcommand{\kms}   {\mbox{{\rm km s$^{-1}$}}}

\newcommand{\msx}{{\it MSX}}

\newcommand{\um}{$\mu$m}
\newcommand{\hii}{\mbox{\ion{H}{2}~}}

\newcommand{\degree}{^{\circ}}
\newcommand{\Msun}{M$_{\sun}$}
\newcommand{\Lsun}{L$_{\sun}$}

\slugcomment{Accepted for publication in the Astrophysical Journal (Letters) 2010 April 9}

\shorttitle{Star Formation in the M17 Proto-OB Association}
\shortauthors{Povich \& Whitney}

\begin{document}


\title{Evidence for Delayed Massive Star Formation in the M17 Proto-OB
Association}


\author{Matthew S. Povich\altaffilmark{1}}
\affil{Department of Astronomy and Astrophysics, The Pennsylvania
  State University, 525 Davey Laboratory, University Park, PA 16802; povich@astro.psu.edu}

\and

\author{Barbara A. Whitney}
\affil{Space Science Institute, 3100 Marine Street, Suite A353, Boulder, CO 80303; bwhitney@spacescience.org}

\altaffiltext{1}{NSF Astronomy and Astrophysics Postdoctoral Fellow}


%\altaffiltext{1}{Visiting Astronomer, Cerro Tololo Inter-American Observatory.
%CTIO is operated by AURA, Inc.\ under contract to the National Science
%\altaffiltext{2}{Society of Fellows, Harvard University.}
%\altaffiltext{3}{present address: Center for Astrophysics,
%    60 Garden Street, Cambridge, MA 02138}
%\altaffiltext{4}{Visiting Programmer, Space Telescope Science Institute}
%\altaffiltext{5}{Patron, Alonso's Bar and Grill}


\begin{abstract}
Through analysis of archival images and photometry from the {\it Spitzer}
GLIMPSE and MIPSGAL surveys combined with 2MASS and \msx\ data,
we have identified 488 candidate young stellar objects (YSOs) in the
giant molecular cloud M17 SWex, which extends ${\sim}50$~pc
southwest from 
the prominent Galactic \hii region M17. Our sample includes ${>}200$
YSOs with masses ${>}3$~\Msun\ that will become B-type stars on the
main sequence. Extrapolating over the stellar initial mass function
(IMF), we find that M17 SWex contains ${>}1.3\times10^4$ young stars,
representing a proto-OB association. The YSO mass function is
significantly steeper than the Salpeter IMF, and early O stars are
conspicuously absent from M17 SWex. 
%Revision 
Assuming M17 SWex will form
an OB association with a Salpeter IMF, %
these results reveal 
the combined effects of (1) more rapid circumstellar disk evolution in
more massive YSOs and (2) delayed onset of massive star
formation. % in the cloud.
\end{abstract}


%% subject header.  Objects should be in the appropriate "individual"
%% headers (e.g. quasars: individual, stars: individual, etc.) with the
%% additional provision that the total number of headers, including each
%% individual object, not exceed six.  The \objectname{} macro, and its
%% alias \object{}, is used to mark each object.  The macro takes the object
%% name as its primary argument.  This name will appear in the paper
%% and serve as the link's anchor in the electronic edition if the name
%% is recognized by the data centers.  The macro also takes an optional
%% argument in parentheses in cases where the data center identification
%% differs from what is to be printed in the paper.

%infrared: ISM --- infrared: stars --- shock waves --- stars:
%  --- HII regions: individual(\objectname{RCW49, M17})}


\section{Introduction}

%More than 3 decades ago, 
\citet{EL76} discovered an extended molecular
cloud associated with the well-known Galactic \hii region M17.
%Discovery of M17 extended GMC by Elmegreen \& Lada
%(1976). 
The size ($70~{\rm pc}\times 15$~pc at $d=2.1$~kpc), mass
\citep[${>}2\times 10^5$~\Msun;][]{ELD79}, and fragmentary morphology
of this cloud were reminiscent of the largest Galactic OB
associations. \citet{EL76} suggested that an extended OB association
will eventually form near M17, and they
%Suggestion that this cloud was the precurser to an OB
%association -- mass and morphology supported this idea. 
noted an apparent progression of ages among the known OB
populations, \hii regions, and molecular cloud cores 
across the complex. %, parallel to the Galactic midplane. 
%Along with Orion and W3, the
The M17 complex became an observational touchstone for
the theory of sequential triggered massive star formation by
propagating ionization fronts \citep{EL77}.

While the M17 \hii region itself and the interface between the
ionization front and the massive molecular core known as M17 SW have
remained the focus of intense study \citep[][and references
therein]{CH08}, interest in the extended molecular cloud, which we
name M17 SWex (since M17 SW is ambiguous), has gradually waned.
% over the
%years.    
%This may be due to the dearth of truly massive stars in 
M17 SWex lacks very massive stars; a
handful of low-luminosity compact and ultracompact \hii regions have
been identified, % consistent with clusters of B stars or late O stars
but they are insufficiently powerful to trigger massive star
formation throughout the cloud \citep{ELD79,JF82}. 
%Focus remained on M17 H II region itself, particularly the
%interaction between the I-front and the most massive cloud immediately
%to the South (the region of peak density is often called M17 SW).

M17 SWex presents one of the most
striking infrared dark cloud (IRDC) morphologies in the Galaxy,
revealed in great detail by the {\it Spitzer Space Telescope} as part
of the Galactic Legacy Infrared Mid-Plane Survey Extraordinaire
\citep[GLIMPSE;][]{GLIMPSE} and the Multiband Imaging Photometer for
{\it Spitzer} Galactic Plane Survey \citep[MIPSGAL;][]{MIPSGAL}.
%large because of relative
%proximity 
%and
%dark because of
%very high column density (numbers?). 
IRDCs have been the subject of much recent work because they are %the
%most likely sites of 
%incipient massive (OB stars, $m\ge 8$~\Msun) star formation in the 
%Galaxy, offering 
laboratories for studying the initial
conditions for massive (OB stars, $m\ge 8$~\Msun) star formation \citep{RJS06,BS07,BT09}.
%({\bf
%  Citations:  Rathborne et al. 2009? papers by Jackson, Rathborne, Chambers, Simon; also papers by Beuther }).  

In this Letter, we use archival GLIMPSE and MIPSGAL data to find and
characterize individual young stellar objects (YSOs) throughout the
${\sim}1\degree$ extent of 
M17 SWex. We model the mass function of the YSO population and find
that it bears the imprints of both mass-dependent circumstellar disk evolution
and delayed massive star formation in M17 SWex. 
%In \S2 we describe our methods for selecting YSOs and
%modeling their physical parameters. In \S3 we present the ensemble
%properties of the YSO population. %, including the spatial distribution
%%and mass function. 
%An analysis of the YSO mass function is
%discussed in \S4, and we summarize our results in \S5.

\section{Data Analysis and Modeling}

\subsection{Selection of Candidate YSOs By SED Fitting of GLIMPSE
  Catalog Sources}

%Dusty circumstellar disks and infalling envelopes surrounding YSOs
%reprocess radiation from the central star, producing 
%characteristic infrared (IR) excess emission. 
%YSOs can be identified
%%in broad-band photometric imaging observations via their IR colors or
% via their 
The infrared (IR) spectral energy distributions (SEDs) of YSOs are dominated
 by emission from dusty circumstellar disks and infalling envelopes. 
For details on our procedure for identifying 
candidate YSOs we refer the reader to
\citet[][hereafter P09]{paper1} and \citet{spitzcar}. 
Here we summarize the main analysis steps. 
%We began with 159,000 sources in the
%highly-reliable GLIMPSE Point Source Catalog \citep{GLIMPSE} 
Using
a $\chi^2$-minimization SED fitting tool \citep{fitter}, we fit
reddened \citet{Kurucz} 
stellar atmospheres (varying $A_V$ from 0 to 40~mag) to the SEDs of
64,820 sources in the 
highly-reliable GLIMPSE Point Source Catalog located
within a $1\degree\times0\fdg75$ field encompassing M17 SWex. 
Only sources detected in
$N_{\rm data}\ge 4$
of the 7 Catalog bands
($\lambda=1$--8~\um) were fit.  
%{\bf (you sort of defined Ndata but may have to define chisqmin)}.
%\citep[$\lambda=1$--8~\um;][]{2MASS,IRAC} were fit.
%, allowing the
%interstellar reddening to 
%vary from $A_V=0$ to 40 mag. 
The best-fit stellar
atmosphere model was a poor fit
($\chi_{\min}^2/N_{\rm data}> 4$; $\chi_{\min}^2$ is the
goodness-of-fit parameter) for 1498 sources; 
%the SEDs
%of these sources are inconsistent with normally reddened stellar
%photospheres, 
hence they are possible YSOs.

We then filtered out sources
with IR excess emission appearing in only the IRAC [5.8] or [8.0]
bands using the \citet{spitzcar} color
criteria, modified to de-redden the $[3.6]-[4.5]$ colors of background stars
viewed through the M17 SWex cloud \citep[using the extinction law of][]{I05}.
This step was necessary to distinguish intrinsically red objects from
sources affected by systematic photometric errors; IRAC [5.8] and
[8.0] are less sensitive and more affected by nebular
emission compared to IRAC [3.6] and [4.5]. We discarded 592
sources, leaving 906 candidate YSOs in our initial sample.

\subsection{Constraints on SED Models from MIPS 24 \um\ Aperture Photometry}

We fit YSO model SEDs from \citet[][hereafter RW06]{grid} to the
available IR photometry 
of each candidate YSO in the initial sample to constrain luminosity, mass,
and evolutionary stage (Stage 0/I, dominated by infalling envelope;
Stage II, optically thick circumstellar disk; Stage III, optically
thin disk; or Ambiguous; see P09). %Because embedded YSOs emit
%primarily in the thermal IR, photometry 
Photometry at 
$\lambda > 10$~\um\ is required to constrain bolometric
luminosities and often to distinguish Stage 0/I from
Stage II sources \citep[RW06;][]{I07}. 
%We performed aperture
%photometry on a MIPSGAL 24~\um\ enhanced mosaic \citep{MIPSGAL}.
%Hence before fitting YSO models to the data, at 
We located the position of each
YSO candidate %in the initial sample
%, we centered 
%an extraction aperture %of radius 3.5\arcsec\ 
%and background annulus
%of inner and outer radii 7\arcsec\ and 13\arcsec\ 
in a MIPSGAL 24~\um\ enhanced mosaic \citep{MIPSGAL} %. We then
and extracted fluxes using aperture photometry.  
%to measure 24~\um\ flux densities for each source, 
%estimating the
%background level using the Daophot MMM algorithm ({\bf STETSON 1987???}).
% CHECK CITATION
%To check the reliability of our aperture photometry, 
We compared the
aperture photometry results to fluxes extracted via PSF 
fitting with the GLIMPSE pipeline for
% same sources 
%found and 
a subset of sources
%We varied the radii of the
%extraction aperture and background radius, and found
%the 
%The best agreement with the PSF fitting results occurred 
and found excellent agreement for aperture radius
3.5\arcsec\ with background annulus 7\arcsec\ to 13\arcsec.
This choice of aperture required an aperture
correction
of 2.8.\footnote{MIPS Instrument Handbook v1.0, \\ \url{http://ssc.spitzer.caltech.edu/mips/mipsinstrumenthandbook/}}
%but
%the uncertainty introduced by the aperture correction is more
%than offset by the greater accuracy of the local background
%determination and the ability to separate close sources. 

Aperture photometry detected more 24~\um\ sources than PSF fitting.
% and
%provided a straighforward means of placing $5\sigma$ upper limits on
%non-detections.
%, a useful constraint for the SED fitting. 
%Taking
%$\sigma$ as the measurement uncertainty, 
For the $35\%$ of
sources lacking 24~\um\ detections we placed $5\sigma$ upper limits,
a useful constraint for the SED fitting.
% the extracted flux density was $F^{\prime}_{24}<5\sigma$, and we set
%upper limits of $F_{24}<5\sigma+F^{\prime}_{24}$ for
%$F^{\prime}_{24}>0$ and $F_{24}<5\sigma$ for
%$F^{\prime}_{24}\le 0$. 
The brightest ${\sim}1\%$ of sources
were saturated in the MIPSGAL images; for these we treated the extracted
fluxes as lower limits.
%$F_{24}>F^{\prime}_{24}$. %, a lower limit for the SED fitting.
For the 24~\um\ detections we set the minimum
uncertainty to 15\% to avoid overly weighting the
24~\um\ flux in the model fits in case of residual systematic
uncertainties. The 
results of our aperture photometry are presented in Table~\ref{table}.
One potential source of systematics is the mid-IR extinction law for
$\lambda>10$~\um. Recent work has found similar reddening at both
24~\um\ and
8~\um\ in dense molecular clouds
\citep[e.g.][]{KF07}. We used the high-density ($A_K\ge 1$) mid-IR
extinction law of 
\citet{MM09} when fitting the
RW06 models to the SEDs of our initial YSO sample.

\subsection{\msx\ Detections of Luminous Candidate YSOs}

%The most luminous members of our YSO sample are bright enough at
%8--21~\um\ to have been included in the \msx\ Galactic Plane Survey
%Point Source Catalog \citep{MSX}. %right name??
We incorporate 8--21~\um\ photometry from the \msx\ Galactic Plane
Survey Point Source Catalog \citep{MSX} into our YSO
sample in 2 ways: (1) Spatial correlation of \msx\ sources with the initial
YSO sample and (2) visual identification of bright sources located near the
IRDCs in the GLIMPSE and MIPSGAL mosaics that were excluded from the
the GLIMPSE Catalog due to saturation. Moderately saturated sources
were included in the more complete GLIMPSE Point Source Archive, and we
added 15 \msx\ sources with Archive matches to the initial YSO
sample. The ${\sim}3\%$ of the YSO sample with \msx\
counterparts (indicated in Table~\ref{table}),  include the most
luminous %, and therefore massive, 
YSOs in the cloud.
The \msx\ data provide strong constraints on the mid-IR 
SEDs along with replacements for {\it Spitzer} measurements
suffering from saturation at 8.0 or 24~\um. 
%{\it Spitzer} bands
%showing signs of saturation or excess emission 
%from PAHs were either excluded from or set as upper/lower limits for
%the SED fitting. 
 
\subsection{Removal of Contaminants from the YSO Sample}

%The initial candidate YSO sample contains sources with SEDs
%that depart from normally reddened stellar photospheres in ${\ge}2$ of
%8 combined 2MASS and {\it Spitzer} bands (up to 12 combined bands for
%sources with \msx\ detections). 
In addition to YSOs, the initial sample may
contain the following classes of contaminants: variable
stars, dusty asymptotic giant branch (AGB) stars, unresolved planetary
nebulae, and background galaxies. The SEDs of variable stars and many
AGB stars 
cannot be fit by stellar atmospheres or YSO models; 15\% of the
initial sample were discarded for this reason. % are not well-fit by the RW06 models and discarded. 
Of the remaining contaminants, luminous AGB stars are 
the most important, as they can
masquerade as massive YSOs (P09). The majority of AGB stars have
$[8.0]-[24]<2.2$~mag
%, and the few extreme AGB stars that are redder
%would be conspicuously high-luminosity \citep{BW08}, we do not find
%any such objects in our sample.
\citep{BW08}, and applying this color cut removes an additional 5\% of
sources. %, leaving 749 candidate YSOs in our sample.

While we expect YSOs associated with M17 SWex to be clustered,
% near the
%molecular cloud fragments, 
contaminating sources, including foreground YSOs, will be distributed
uniformly. Following P09, we used  
a cluster-finding algorithm to select sources exhibiting a
significant degree of clustering with respect to a 
%uniformly
%distributed 
``control'' region located away from the molecular
cloud. 
%< Revisions
The control field sampled was a $0\fdg38\times 0\fdg35$
box centered at $(l,b)=(13\fdg89,-0\fdg76)$ with source
density 270 deg$^{-2}$, yielding a similar average source separation
($\Theta_{\rm con}=3.84\arcmin$)
to the P09 control field. 
The surface density ratio between the control field and the entire
field is $\Sigma_{\rm con}/\Sigma_{\rm targ}=0.27$,
which shows that M17 SWex presents a significantly larger
overdensity of candidate YSOs versus contaminants compared to the M17
molecular cloud sample analyzed by P09.
%>
%The average density of 
%1000 deg$^{-2}$ over the entire % $1\degree\times 0\fdg75$
%field. 
%The algorithm was tuned to be most sensitive to groupings of 5
%candidate YSOs. 
We thereby split the initial sample into a clustered
population (${\sim}65\%$) comprised of YSOs associated with M17 SWex
and a distributed component (${\sim}35\%$) dominated by unassociated
contaminants. 
 
\section{Results}

\subsection{Final YSO Sample}

%%%%% FIGURE 1 %%%%%
%
%\begin{figure}
\begin{figure*}
%\epsscale{1.25}
\epsscale{0.9}
\plotone{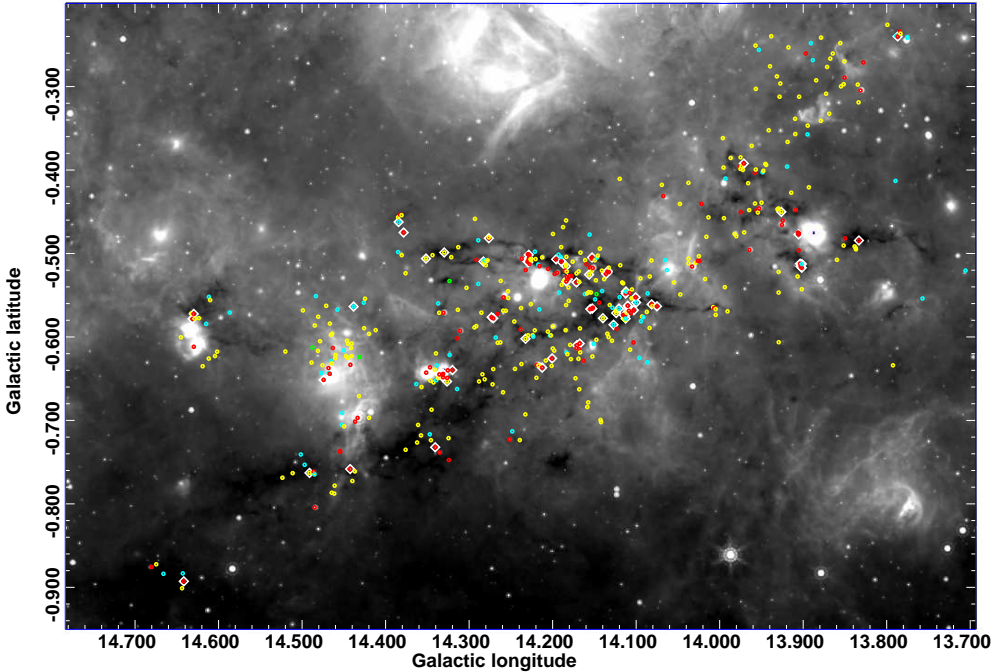}
\caption{MIPSGAL 24~\um\ mosaic image with positions of the 488 YSOs
  in the final sample overlaid. Colors identify YSO evolutionary
  stages: {\it red}=Stage 0/I, {\it yellow}=Stage II, {\it green}=Stage III, {\it cyan}=ambiguous. YSOs with 4.5~\um\ excess
  emission are highlighted with {\it white} diamonds.
\label{image}}
%\end{figure}
\end{figure*}
%
%%%%%%%%%% TABLE 1 %%%%%%%%%
%
\begin{deluxetable*}{rcc@{  }c@{  }ccccccc@{  }c@{  }c}
%\begin{deluxetable}{rcccccccccccc}
%\rotate
\tabletypesize{\footnotesize}
%\begin{table*}
%\caption{Basic Properties of Candidate YSOs\label{table}}
\tablecaption{Basic Properties of Candidate YSOs\label{table}}
\tablewidth{\textwidth}
%\tablewidth{0pt}
\tablehead{
%\colhead{Cat.} & \colhead{GLMA} & \colhead{[24]\tablenotemark{a}} & \colhead{}
%& \colhead{$\langle\log{L_{\rm bol}}\rangle$} & \colhead{$\sigma(\log{L_{\rm bol}})$} & \colhead{$\langle M_{\star}\rangle$} & \colhead{$\sigma(M_{\star})$} & 
%\colhead{$\langle\log{\dot{M}_{\rm env}}\rangle$\tablenotemark{c}} & \colhead{$\sigma(\log{\dot{M}_{\rm env}})$}
% & & & \colhead{4.5~\um}\\
%\colhead{No.} & \colhead{Name} & \colhead{(mag)} &
%\colhead{$q_{24}$\tablenotemark{b}} &
%\colhead{(\Lsun)} & \colhead{(\Lsun)} & \colhead{(\Msun)} &
%\colhead{(\Msun)} & \colhead{(\Msun~yr$^{-1}$)} & \colhead{(\Msun~yr$^{-1}$)} & 
%\colhead{Stage} & 
%\colhead{\it MSX?} & \colhead{Excess?}
%}
%\begin{tabular*}{\textwidth}{rcccccccccc@{  }c@{  }c}
%\hline\hline\\[-5pt]
{Cat.} & {GLMA} & {[24]\tablenotemark{a}} & {}
& {$\langle\log{L_{\rm bol}}\rangle$} & {$\sigma(\log{L_{\rm bol}})$} & {$\langle M_{\star}\rangle$} & {$\sigma(M_{\star})$} & 
{$\langle\log{\dot{M}_{\rm env}}\rangle$\tablenotemark{c}} & {$\sigma(\log{\dot{M}_{\rm env}})$}
 & & & {4.5~\um~} \\
{No.} & {Name} & {(mag)} &
{$q_{24}$\tablenotemark{b}} &
{(\Lsun)} & {(\Lsun)} & {(\Msun)} &
{(\Msun)} & {(\Msun~yr$^{-1}$)} & {(\Msun~yr$^{-1}$)} & 
{Stage} & 
{\it MSX?} & {Excess?} 
}
%\hline
\startdata
 200 & G014.1482-00.5321 &  7.7 & 1 &  0.7 &  0.3 &  1.3 &  0.6 &  -5.9 &   2.6 &  II &            &            \\
 201 & G014.1491-00.5550 &  6.1 & 1 &  1.5 &  0.2 &  2.6 &  0.9 &  -5.2 &   2.4 &   A &            &            \\
 202 & G014.1494-00.5081 &  6.4 & 3 &  2.1 &  1.2 &  3.5 &  0.9 &  -5.7 &   2.4 &  II &            &            \\
 203 & G014.1508-00.6084 &  3.3 & 1 &  2.7 &  0.7 &  3.9 &  2.5 &  -4.2 &   1.7 &   A &            &            \\
 204 & G014.1519-00.5650 &  6.0 & 1 &  1.1 &  0.8 &  1.1 &  1.0 &  -4.7 &   2.1 & 0/I &            &            \\
 205 & G014.1520-00.5173 &  0.6 & 2 &  3.6 &  0.4 &  8.4 &  1.5 &  -3.2 &   1.5 & 0/I & \checkmark &            \\
 206 & G014.1524-00.6234 &  6.1 & 1 &  1.8 &  0.2 &  2.8 &  0.3 &  -7.2 &   2.9 &  II &            &            \\
 207 & G014.1525-00.5661 &  6.1 & 1 &  1.7 &  1.1 &  1.2 &  0.9 &  -3.7 &   0.6 & 0/I &            & \checkmark \\
 208 & G014.1528-00.5048 &  5.8 & 1 &  1.5 &  0.9 &  1.6 &  1.5 &  -4.0 &   1.8 & 0/I &            & \checkmark \\
 209 & G014.1530-00.5209 &  5.3 & 1 &  1.8 &  0.2 &  2.8 &  0.6 &  -5.5 &   2.4 &  II &            &            \\
 210 & G014.1532-00.5061 &  6.1 & 3 &  1.5 &  0.9 &  2.3 &  0.4 &  -4.0 &   0.1 & 0/I &            &            \\
 211 & G014.1542-00.5467 &  7.3 & 1 &  1.1 &  0.7 &  1.9 &  1.0 &  -5.9 &   2.6 &  II &            &            \\
 212 & G014.1549-00.5666 &  4.7 & 1 &  1.5 &  0.9 &  1.0 &  1.1 &  -4.3 &   1.8 & 0/I &            & \checkmark \\
 213 & G014.1552-00.5169 &  3.6 & 1 &  1.9 &  1.1 &  2.8 &  1.4 &  -4.1 &   1.9 & 0/I &            &            \\
 214 & G014.1557-00.4774 &  5.8 & 1 &  1.6 &  1.0 &  2.2 &  1.1 &  -4.8 &   2.1 &   A &            &         
\enddata
%\hline
%\end{tabular*}
\tablecomments{Table 1 is available in its entirety in the electronic
  edition of the {\it Astrophysical Journal Letters}. A portion is
  reproduced here for guidance regarding its form and content.}
\tablenotetext{a}{Typical uncertainty on [24] is ${\sim}0.1$~mag.}
\tablenotetext{b}{24~\um\ quality flag: 0=non-extraction, 1=detection,
2=lower limit, 3=upper limit.}
\tablenotetext{c}{If $\langle \dot{M}_{\rm
    env}\rangle<10^{-9}$~\Msun~yr$^{-1}$, accretion has effectively
  ceased, hence $\langle\log{\dot{M}_{\rm env}}\rangle$ is undefined.}

%\end{table*}
\end{deluxetable*}
%\end{deluxetable}
%
The clustered population of 488 sources comprises our final
YSO sample. % (Fig.~\ref{image} \& Table~\ref{table}). 
The YSOs exhibit a highly structured spatial distribution, strongly
clustered along the IRDC filaments with numerous sub-clusters
(Fig.~\ref{image}). Using the set $i$ of well-fit models for each
YSO (defined by $\chi^2_i-\chi^2_{\min}\le 2N_{\rm data}$), we
construct probability distributions of bolometric luminosity
$L_{\rm bol}$, mass of central star $M_{\star}$, and evolutionary stage
(see P09 for details). We allowed the model fitting tool to accept a
range of distances from 1.9 to 2.3~kpc, hence incorporating
the uncertain distance to the M17 complex into the uncertainties on
the resultant model parameters (P09).
The results for each source are summarized in
Table~\ref{table}. The final sample includes 133 Stage 0/I, 276
Stage II, 4 Stage III, and 75 YSOs with ambiguous stage. The paucity of
Stage III objects is a selection effect and suggests that virtually
all of the ambiguous sources are ambiguous between Stages 0/I and
II. Stage 0/I objects are more tightly clustered 
%in the IRDCs 
than are Stage II, % YSOs % and 2 looser groups of sources (near
%$[l,b]=[14\fdg3,-0\fdg6]$ and $[13\fdg9,-0\fdg6]$) dominated by Stage
%II 
supporting the idea that the stages represent an evolutionary
sequence. The high fraction of Stage 0/I sources ($N_{\rm
  0I}/N_{\rm II}\approx 0.5$) confirms the extreme youth of the population.

%Another likely indicator of youth, 
Among the final sample, 68 YSOs
exhibit excess 4.5~\um\ emission over the best-fit YSO model.
These may be unresolved analogs of the ``extended green objects'' 
(EGOs), candidate accretion-powered young massive stellar outflows
 \citep{egos}. The RW06 models do not
include outflows, which can produce strong molecular line emission 
in the 4.5~\um\ band. 
We set the 4.5~\um\ fluxes to upper
limits for the SED fitting in these cases. 
%The 4.5~\um\ excess sources
%are indicated in 
%Table~\ref{table} and Fig.~\ref{image}. 
The majority of 4.5~\um\ excess sources are
associated with tight clusters of Stage 0/I YSOs
(Fig.~\ref{image} and Table~\ref{table}). Two such groups coincide
with known EGOs 
\citep[G014.33-0.64 and G014.63-0.58;][]{egos}.

\subsection{Observed YSO Mass Function}

Given the GLIMPSE sensitivity limits and the 2.1~kpc distance to M17,
our YSO sample is nearly complete for $m=M_{\star}\ga 3$~\Msun,
corresponding to main-sequence spectral types earlier than A0 (P09). For this
mass range, the pre-main-sequence evolutionary tracks are horizontal
\citep{BM96,SDF00}, and YSOs follow the zero-age main
sequence mass--luminosity relation %: $L_{\rm bol}\approx L_{\rm
                                %ZAMS}$ 
\citep{MK09}.

%%%%% FIGURE 2 %%%%%
%
\begin{figure}
\epsscale{1.0}
\plotone{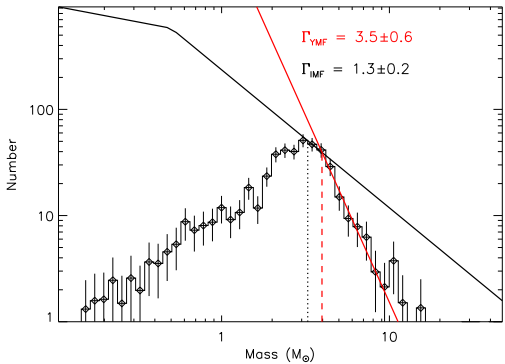}
\caption{YMF plot (histogram points with error bars). Two IMFs are
  overplotted: power-law fit to the high-mass tail of the YMF
  ($m\ga 4$~\Msun; {\it red}) and \citet{Kroupa} IMF scaled to
  match the YMF peak at ${\sim}3.2$~\Msun\ (dotted
  line). The IMFs intersect at 
  $M_c=3.9$~\Msun\ (dashed line).
\label{YMF}}
\end{figure}

Summing the probability distributions of $M_{\star}$ from the model fits
to each YSO, we construct the YSO mass function (YMF; P09) for the 488
sources in our final sample (Fig.~\ref{YMF}). The YMF has the same
form as the stellar initial mass function (IMF),
\begin{equation}\label{IMF}
\Phi(\log m)={dN}/{d\log m}\propto m^{-\Gamma},
\end{equation}
where $dN$ is the number of stars in the (logarithmic) mass interval
$(\log m,~\log m + d\log m)$ and $\Gamma$ is the power-law ``slope''
\citep{BCM10}. %The high-mass ($m>0.5$~\Msun) slope of 
Measurements of the IMF over a wide variety of environments, from
young open clusters to field stars,  
have found the slope to be remarkably consistent with the
\citet{Salpeter} IMF, $\Gamma_{\rm IMF}=1.3\pm0.2$ for $m>0.5$~\Msun\
\citep{Kroupa}. The YMF slope, however, is significantly
steeper; $\Gamma_{\rm YMF} = 3.5\pm 0.6$ above a critical mass, $m\ga
M_c=3.9$~\Msun.
The YMF apparently flattens to $\Gamma_{\rm IMF}$ for 3.2~\Msun~$\la m
\la M_c$ before 
turning over due to incompleteness at lower masses
(Fig.~\ref{YMF}). Such a steep slope translates into a 
glaring deficit of massive stars; the YMF contains ${>}200$ YSOs equivalent to
main-sequence B stars ($m>3$~\Msun) but {\it zero} O stars
($m>20$~\Msun), and ${\sim}80$ OB stars with $m\ge M_c$ are missing
compared to the predictions of a Salpeter--Kroupa IMF.

\section{Discussion}

\subsection{YSO Mass Spectrum Parameterization}

%<Added in revision -
Assuming a Salpeter--Kroupa IMF applies to M17 SWex, % >
the steep YMF
slope can be understood in 
terms of evolution and 
selection effects. Our sample %(1) 
is populated by
ongoing star formation and %(2) 
includes {\it only} YSOs
with circumstellar material (disks). We parameterize these effects
using the linear form of the  
mass function, the {\it mass spectrum,}
\begin{equation}
  \phi(m)={dN}/{dm}\propto m^{-\alpha},~~\alpha=\Gamma+1
\end{equation}
\citep{BCM10}. The YSO mass spectrum $\psi(m)$ is populated over time as
\begin{equation}\label{prime}
  \psi^{\prime}(m)=\int_{\tau_0(m)}^{\tau}\frac{\partial\phi}{\partial{t}}dt,
\end{equation}
where $\tau$ is the time since star formation began in the cloud,
$\partial\phi(m)/\partial{t}$ is the star formation rate (SFR) in 
mass interval $(m,~m+dm)$, and $\tau_0(m)$ is the time at which star
formation began {\it for each mass interval}. 
%This allows for a delay
%in the onset of massive star formation, for example. 
%({\bf do we need
%  some rationale/citations here?}). %CITATIONS?
%Over the mass range of
%the {\it first} stars to begin forming, $\tau_0\equiv 0$, but for $\tau_0(m)>\tau$
Assuming for simplicity that the SFR is not time-dependent,
$\partial\phi(m)/\partial{t}\to\phi(m)/\tau$ and the integral in
Equation~\ref{prime} becomes
\begin{equation}\tag{3a}\label{primea}
  \psi^{\prime}(m)=\phi(m)[1-\tau_0(m)/\tau]=\phi(m)f_0(m),
\end{equation}
where $f_0(m)$ is the fraction of stars that have {\it already} formed
versus all stars that will {\it eventually} form in each mass interval.
In the limiting cases of $\tau_0(m)=0$ (first stars to
form in cloud) or $\tau_0(m)\ll\tau$ (timescale for differential onset
of star formation small compared to age of population),
$f_0(m)=1$ and $\psi^{\prime}(m)=\phi(m)$.

The YSO sample is further biased by disk evolution. Taking $\tau_d(m)$
to be the disk lifetime as a function of mass,
% and assuming disk
%destruction rates in each mass bin are not time-dependent, 
\begin{equation}\label{disks}
%  \psi(m)=\psi^{\prime}(m)[{\tau_d(m)}/{\tau}]=\psi^{\prime}(m)f_d(m),
  \psi(m)=\psi^{\prime}(m)\frac{\tau_d(m)}{\tau-\tau_0(m)}=\psi^{\prime}(m)f_d(m),
\end{equation}
where $f_d(m)$ is the (mass-dependent) disk fraction. For
$\tau_d(m)\ge\tau-\tau_0(m)$, $f_d(m)=1$.

Combining
Equations \ref{primea} and \ref{disks},
\begin{equation}\label{combined}
  \psi(m)=\phi(m)f_0(m)f_d(m). %\propto m^{-\alpha}m^{-\omega}m^{-\delta}
\end{equation}
Since the observed YMF can be fit with a power law (Fig.~\ref{YMF}),
we parametrize $f_0(m)$ and $f_d(m)$ as $m^{-\omega}$ and
$m^{-\delta}$, respectively. Hence the observed deviation of the YMF
slope from the standard IMF is
\begin{equation}\label{powers}
  \Delta\alpha=\Delta\Gamma=\Gamma_{\rm YMF}-\Gamma_{\rm IMF}=\omega+\delta=2.2\pm 0.6
\end{equation}
for $m\ge M_c$ (Fig.~\ref{YMF}).

\subsection{Implications of a Steepened YSO Mass Function: Disk Evolution,
  Delayed Massive Star Formation, or Both?}

We have shown that a steep YMF could be produced by
delayed formation of massive stars or by shorter disk lifetimes for
more massive stars. Both interpretations have physical basis. 
%{\bf Matt:  A
%  couple sentences here citing theoretical references for each mechanism?}
%  {\bf, Barb: I don't think it's needed}.
Can either of these effects alone explain the observed YMF?

%%%%% FIGURE 3 %%%%%%
%
\begin{figure}
\epsscale{1.0}
\plotone{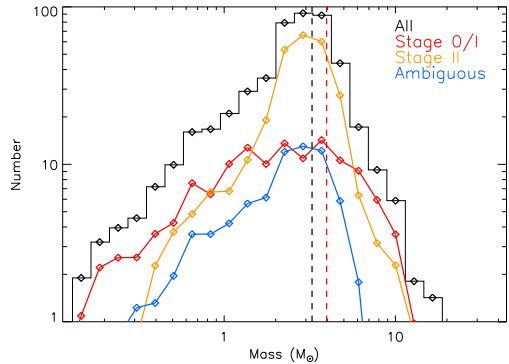}
\caption{YMF of Fig.~\ref{YMF}, binned up by a factor of 2 and
  subdivided by evolutionary stage. 
\label{YMF_stages}}
\end{figure}
{\it Disk evolution.} The YMF is broken down by evolutionary stage in
Fig.~\ref{YMF_stages}, and it is apparent that the steep slope is
driven by Stage II sources. %, which dominate the sample. This motivates
%us to 
We thus consider the case of disk evolution alone:
$\omega\to0$, $\delta\to2.2$ (Equation~\ref{powers}). \citet{JH07} measured
the disk fractions in the $\sigma$ Orionis cluster (age ${\sim}3$~Myr)
to be ${\sim}35\%$ for $m\le1$~\Msun\ and ${\sim}10\%$ for
$m>2$~\Msun, giving $\delta<1.8$. 
%{\bf Do you want to mention the possibility that there is a sharp cutoff at high mass due 
%to the intense radiation fields?  i.e., that it's not necessarily a simple power law?}
This does not
rule out disk evolution as the primary driver of the YMF slope, given
the uncertainties, but the disk lifetimes themselves, which set
the duration of the Stage II phase, become 
problematically short. Adopting $\delta = 1.8$ and $\tau_d(1$~\Msun$)=2$~Myr
\citep{HLL01}, we would predict $\tau_d(4$~\Msun$)=0.16$~Myr. 
%{\bf Here it might 
%make it clearer to the reader if you remind them that taud is also the Stage II lifetime.
%Both terms are being used to mean the same thing I think.} 
This is
comparable to the typical duration of the %Class I 
envelope-accretion phase in low-mass
stars \citep[${\sim}0.1$~Myr;][]{KH95}. %{\bf or Whitney \& Hartmann?}
The Stage 0/I objects in our sample span a wide range in mass
(Fig.~\ref{YMF_stages}), yet the typical accretion timescale is similarly
$\tau_A\sim0.1$~Myr
($\tau_A=M_{\star}/\dot{M}_{\rm env}$, where $\dot{M}_{\rm env}$ is the
accretion rate in the RW06 models; see Table~\ref{table} and P09).
%Stating the problem in reverse, we 
We expect the {\it maximum}
age of a 4~\Msun\ Stage II source in our sample to be $(N_{\rm
  II}/N_{\rm 0I}+1)\tau_A\sim0.7$~Myr ($N_{\rm II}/N_{\rm 0I}\approx 6$
at $m=4$~\Msun; Fig.~\ref{YMF_stages}). This longer, more realistic
disk lifetime implies $\delta\sim 0.9$. 

{\it Delayed Massive Star Formation.} 
%Since disk evolution cannot
%account for all of the steepening in the YMF slope, we 
%The YMF therefore indicates that massive star
%formation has been delayed in M17 SWex ($\omega\ne 0$). 
%Additional
%lines of evidence support this interpretation. 
%We therefore conclude that delayed 
Since disk evolution alone cannot explain the
observed YMF, delayed massive star formation must {\it also} 
contribute ($\omega\ne 0$).
While diskless
low- and intermediate-mass stars are lost among the overwhelming
field star 
population in the IR images, %of the inner Galactic plane, 
massive stars cannot easily hide within a dense molecular cloud. The
\citet{Kroupa} 
IMF, scaled to match the YMF peak (Fig.~\ref{YMF}), predicts ${>}10$ O
stars, including at least one early O star
($m>50$~\Msun); such stars ionize dusty compact and ultracompact
\hii regions. The few \hii regions in M17 SWex (visible as compact,
bright extended sources in Fig.~\ref{image}) are insufficiently
luminous to contain O stars \citep{ELD79,JF82}. We note also that the
ratio $N_{\rm II}/N_{\rm 0I}$ steadily decreases with $m$ for $m\ga 4$~\Msun\
and actually inverts for $m>6$~\Msun\ (Fig.~\ref{YMF_stages}),
suggesting that the more massive 
YSOs are preferentially younger. 
%{\bf Worth mentioning here that this is thus a proto-Ostar
%cloud in the making?  you mention it in the conclusions but twice can
%be good}.  
This is not a firm result, however, because
it is based on ${<}20$ sources, and the trend may instead
reflect instability of disks around massive YSOs. 
%(a proper Stage
%II phase may not exist for massive stars).

%We thus conclude that {\it both} disk evolution and delayed massive
%star formation conspire to steepen the YMF slope. 
While the data
presented here do not support more than simple parameterizations, we
can constrain the delay timescale for the most massive stars:
$\tau_0(m>20$~\Msun$)>\tau$. 
If the YMF break at $M_c\sim 4$~\Msun\ (Fig.~\ref{YMF}) is real, 
not an artifact of incompleteness, then $\psi(m\le M_c)=\phi(m\le
M_c)$, hence $f_d(m\le M_c)=1$
(Equation~\ref{combined}), and $\tau=\tau_d(M_c)\sim0.7$~Myr
(Equation~\ref{disks}) is age of the oldest YSOs in the cloud.  
%{\bf, i.e., the age of the cloud?}

\subsection{Present-Day Star Formation Rate}

Integrating the scaled \citet{Kroupa} IMF (Fig.~\ref{YMF}) over $m\ge
0.1$~\Msun\ yields a total population of $1.3\times 10^4$ YSOs in M17
SWex, with a total stellar mass of $8\times 10^3$~\Msun. \citet{EL76}
predicted that M17 SWex would eventually form an OB association; these
numbers, lower limits due to possible incompleteness, show that the
M17 proto-OB association is already 
forming. Adopting $\tau=0.7$~Myr, the present-day SFR in M17 SWex is
0.011~\Msun~yr$^{-1}$, comparable to the time-averaged SFR of M17
itself (P09), albeit distributed over a much larger volume.

\subsection{Sequential Star Formation in the M17 Complex}

\citet{ELD79} compared the morphology of the M17 complex to the dust
lanes and ``beads-on-strings'' \hii regions of
extragalactic spiral structure. 
%The most recent observations
%bolster such a comparison. 
P09 discovered an extended, diffuse \hii region
northeast of M17, called M17 EB, ionized by a group of optically
revealed O stars in a 2--5~Myr old cluster or association. When
M17 SWex is included, the entire M17  
complex presents a clear sequence of star formation extending
${>}100$~pc %from NE to SW, 
parallel to the Galactic midplane and
spanning several Myr in age.
%: (1) faint, diffuse \hii region (M17 EB);
%(2) bright, embedded \hii region (M17); and (3) young proto-OB
%association lacking bright \hii regions (M17 SWex). 
As \citet{EL76}
noted, Galactic spiral density waves propagate in the direction
of decreasing age. % of the major OB populations. 
The difference between the Galactic rotation speed and the spiral pattern
speed at the location of M17 is $221~\kms-130~\kms=91~\kms$
\citep{BB93,MM04}. The timescale for a spiral shock to cross the
%${\sim} 100$~pc 
%extent of the 
entire M17 complex % to pass through the Sagittarius arm 
is ${\sim}1$~Myr, the same order of magnitude as the observed age
spread.

The rapid passage of the M17 complex through the Sagittarius spiral arm
served as the ``global'' trigger ultimately responsible for the
formation of this large OB association. Has sequential
``local'' triggering driven by OB stars \citep{EL77} played an
important role? % Here the picture becomes complicated. 
While there is
strong circumstantial evidence for locally triggered star formation
on the periphery of the M17 and 
M17 EB \hii regions (P09 and references therein), it remains unclear whether any of
the {\it major} OB clusters were similarly triggered.
%, as opposed to
%the triggering of less massive clusters at the periphery of the
%\hii regions. 
Global triggering appears to be a far more likely
explanation for the formation of the proto-OB association in M17 SWex.
%The distribution in space is large ${\sim}50$~pc, while the spread in
%ages in very small ${\la}0.6$~Myr.
The ${\la}0.7$~Myr spread in ages among the YSO sample is small 
%${\la}0.6$~Myr %that the population can be considered coeval 
compared to the 2--3~Myr timescale for sequential triggering by OB stars
\citep{EL77}. %, yet YSOs are distributed spatially across the entire
%${\sim}50$~pc 
%extent of M17 SWex. 
The spatial distribution of both the stars and the dust
(Fig.~\ref{image}) suggest a disturbed molecular cloud that is
experiencing global collapse and fragmentation.
% ({\bf citation?}). {\bf
%Matt:  Is the reasoning in the preceding sentences sound?  Barb:  I think so}
                    
\section{Summary}

We have 
%examined archival catalogs and images from IR Galactic plane
%surveys and 
catalogued 488 predominantly intermediate-mass YSOs in 
%the
M17 SWex. 
%GMC complex. 
Analysis of our YSO sample yields the following main results:
\begin{itemize}
\item M17 SWex will form ${>}200$ B stars. The total
  stellar population in the cloud is ${>}1.3\times 10^4$,
  with total stellar mass ${>}8\times 10^3$~\Msun. The present-day SFR is
  ${\sim}0.011$~\Msun~yr$^{-1}$. M17 SWex is therefore forming
  a proto-OB association. 
\item O stars are
  conspicuously absent. The YMF slope is
  significantly steeper than Salpeter for $m>4$~\Msun. %The steep slope
%  This is explained by the combination of 2 effects: 
\item Disk evolution
  proceeds more rapidly for higher-mass stars, with typical disk
  lifetimes of ${\la}0.6$~Myr for YSOs with $m>4$~\Msun. 
\item Massive stars
  begin forming at later times than low-mass stars. M17 SWex
  probably has not yet formed its most massive star, predicted to be
  an early O-type star of $m>50$~\Msun.
%\item 
\end{itemize}
If a ${>}0.5$ Myr delay in the onset of massive star formation is
the rule in Galactic giant molecular clouds, the best places to study
the initial 
conditions of 
massive star formation may be clouds where intermediate-mass star
formation has already begun.

\acknowledgments
We thank T. P. Robitaille for providing the YSO models and SED
fitter, M. R. Meade for performing 24~\um\ PSF photometry,
L. K. Townsley and 
P. S. Broos for useful conversations, and the anonymous referee for
helpful comments.
M.S.P. is supported by an NSF Astronomy and Astrophysics Postdoctoral
Fellowship under award AST-0901646. B.A.W. acknowledges support from
the Spitzer GLIMPSE programs, Spitzer Theory Program, and NSF AST-0507164.


%% copy editing.  Individual instruments can be provided in parentheses,
%% after the keyword, but they will not be verified.

Facilities: \facility{Spitzer}, \facility{2MASS}, \facility{MSX}



\end{document}